# A Simple Cooperative Platooning Controller for Connected Vehicles

**Youssef Bichiou, Ph.D.**
Research Associate
Center for Sustainable Mobility
Virginia Tech Transportation Institute, Blacksburg, VA 24061
Email: youssef1@vt.edu

**Hesham Rakha, Ph.D., P.Eng. (Corresponding author)**
ORCiD: https://orcid.org/0000-0002-5845-2929
Samuel Reynolds Pritchard Professor of Engineering and Director
Center for Sustainable Mobility
Virginia Tech Transportation Institute, Blacksburg, Virginia 24061
Email: hrakha@vt.edu

**Hossam M. Abdelghaffar, Ph.D.**
Assistant Professor at Mansoura University (Egypt) and Research Associate
Center for Sustainable Mobility
Virginia Tech Transportation Institute, Blacksburg, VA 24061
Email: hossamvt@vt.edu

## ABSTRACT

Efficiency is one of the main quests behind many developed technologies. In the field of transportation, efficiency is usually intertwined with enhanced mobility and reduced costs. This is particularly true for congestion alleviation. Urban traffic congestion is a chronic problem faced by many cities. It is essentially inefficient infrastructure use which results in increased vehicle fuel consumption and emissions. This in turn adds extra costs to commuters and businesses. Addressing this issue is therefore of paramount interest due to the perceived dual benefit. Many technologies were and are being developed. These include adaptive traffic signals, dedicated lanes, etc. This paper presents a simple platooning algorithm that maintains relatively small distances – predetermined time gap – between consecutive vehicles to enhance mobility, increase transportation capacity and ultimately reduce travel costs. Several dynamic and kinematic constraints governing the motion of vehicles are also accounted for. These include acceleration, velocity, and distance constraints. This developed logic was tested on highways that traverse the downtown area of Los Angeles. Depending on the Market Penetration Rate (MPR) of connected automated vehicles (CAVs) versus non-CAVs, a reduction in travel time, delay and fuel consumed across the city can be observed. Vehicles are expected to reach their destination in less travel time, delay and consumed fuel. The reduction percentages range from 0 to 5%, 0 to 9.4%, and 2.58% to 8.17%, respectively. These numbers apply to all the vehicles and not only the ones that form platoons.

# INTRODUCTION

Roadway traffic is a complex phenomenon that requires tedious modeling. Nevertheless, various properties can be directly observed. These properties include (1) traffic stream density ($k$): the number of vehicles per unit length per road or lane; and (2) space-mean speed ($u$): the traffic stream density weighted average speed. Congestion is intertwined with high density and slow space-mean speed. Researchers are attempting to reduce its impact by developing solutions that route traffic and make use of the infrastructure as efficiently as possible. The advancements in wireless communication, and the introduction of various and advanced driver assistance systems [1] allowed several ideas developed on paper to become a reality. One of these ideas is platooning. Platooning – also known as cooperative adaptive cruise control (CACC) – is essentially, a group of vehicles driving at the same speed and maintaining a relatively small distance in-between. This simple concept has the potential to enhance mobility and increase the capacity of the road. The perceived advantages of platooning include efficient mobility, lower fuel consumption, and increased roadway capacity.

Particular attention was and still being allocated to the development of platoons for Heavy Duty Vehicles (HDV). This is primarily motivated by the perceived gains in terms of fuel consumption and as a results significant reduction in $CO_2$ emissions. The work of Deng and Ma [2] used Pontryagin Maximum Principle (PMP) to develop a platooning algorithm for trucks. They claim up to 30% reduction in fuel consumption on the deceleration regime and up to 3.5% in the acceleration regime.

Other nationally funded projects targeting platooning technologies for HDVs were also unveiled. Within the framework of the Japanese national Intelligent Transportation System (ITS) project, Tsugawa et al. [3] developed a technology that could help trucks platoon automatically. They showed that field tests on a three fully automated trucks driving at velocities of 80 km/h and a distance gap of 10 m (i.e. a time gap of 0.45 (s)) can lead to a 14% saving in fuel consumption and thus a reduction in $CO_2$ emission. Carl et al. [4] detailed the various projects that focuses on platooning, namely Safe Road Trains for the Environment (SARTRE) (a European platooning project), Partial Automation for Truck Platooning (PATH) (a California traffic automation program), Grand Cooperative Driving Challenge (GCDC) (cooperative driving initiative), SCANIA Platooning and Energy ITS. He stressed the importance of Vehicle to Vehicle (V2V) and Vehicle to Infrastructure (V2I) (commonly referred to as V2X) communication, and the vehicle and non-vehicle sensors in making these solutions possible in terms of synchronization and vehicle longitudinal and lateral stability.

Davila et al. [5] highlighted the increased safety and reduced environmental impact of platooning. Within the framework of SARTRE project, they performed virtual testing of what platooning can achieve in terms of fuel consumption reduction due to "enhanced" aerodynamics. They pointed out that the increased safety is primarily due to the automated nature of platooning since 95% of the accidents involve human factor contribution [6]. They also noted that the contribution of the aerodynamic drag to fuel consumption overtakes the contribution of the rolling resistance starting from the velocity of 70 km/h. In another study, Michael et al. [7] performed a series of platooning tests on two 'class 8 vehicles' (i.e. a vehicle whose gross vehicle weight rating (GVWR) exceeds 33000 lb) on a closed test track. They varied the speed, the distance gap and mass of the vehicles to determine the best combination leading to the lowest fuel consumption. They reported that for cruising velocity of 88 km/h (55 mph) and a 9.1-m (30-ft) gap distance, the maximum amount of fuel that could be saved is 6.4 %. They also noted that in some occasions the engine coolant fan of the trailing vehicle did not operate. Heavy loads affect the percentage saving; however, the percentage is still significant, they noted. The developed technology would require a modest investment. However, an attractive return on investment is attainable [7]. They also stressed the need to re-design aerodynamic aids for platoons rather than the current vehicle aids in order to further reduce the fuel consumed.

Ellis et al. [8] highlighted the unintended consequences of platooning on the vehicles, particularly heavy trucks. Using computational fluid dynamics, they stressed the significant aerodynamic drag reduction. However, they noticed that if the gap between trucks is small (i.e. 5m), the air flow going through the engine is significantly reduced which results in the fan of all trucks behind the lead being continuously engaged. This according to the authors mitigates the potential fuel economy.

Al Alam et al. [9] acknowledged in their work the inconclusiveness with respect to the fuel consumption reduction for trucks while traveling in platoons. They considered two trucks forming a platoon and equipped with a commercial Adaptive Cruise Control (ACC). Their study concluded that a maximum fuel reduction of 4.7 to 7.7% depending on the time gap could be achieved with maximum savings corresponding to a time gap of 1 (s). They also showed that the ACC does not increase fuel consumption. They recommended a shorter time gap for maximum reduction of drag but at the same time acknowledged the challenges associated, namely, feedback delays and communication delay could endanger the safety and comfort of drivers.

Another work by Vegendla et al. [10] investigated the aerodynamic influence of multiple on-highway trucks in different platooning configurations. Using computational fluid dynamics, they discovered that the configuration with two trucks in one single lane results in an up to 23 % reduction in fuel consumption. They also discovered that two trucks traveling side by side in a two-lane road for instance results in excessive fuel consumption. They reported an 11 % increase in fuel burned by the two vehicles.

Beside the technologies developed for HDVs similar technologies were and are being developed for passenger cars. For instance, Stanger and Del Re [11] developed a linear predictive control model that directly optimizes the fuel consumption of the vehicles inside the platoon. A simplified car-following model was adopted and a quadratic approximation of the fuel consumption was chosen. With this approach they claim a 20% reduction in fuel. Other elaborate models were also proposed. For example, exploiting a non-linear vehicle model Schmied et al. [12] developed a Nonlinear Model Predictive Control (NMPC) logic taking into account various nonlinear constraints. It is important to mention here that the nonlinear nature of the model presents a computational burden preventing its real-time implementation. Nevertheless, the controller was tested using a Hardware-in-the-loop (HIL) configuration and the authors claim a 13% reduction in fuel consumption as well as a 24% reduction in $NO_x$ emissions.

The present effort delivers a simple platooning logic principally inspired by a change of variables. It extends the previous research efforts in the following aspects: (1) it considers platoons of arbitrary lengths; (2) the platoons are formed and broken in a dynamic fashion; and (3) the algorithm is tested on a large-scale virtual implementation. In order to simplify the analysis, this effort only focuses on platoons composed solely of passenger vehicles. Buses or trucks are not considered. In the following section, we enumerate the various forces acting on a moving vehicle as well as the dynamic and kinematic constraints. Details of the controller and the simulation setup are presented in the same section. In the third section, results are discussed and in the last section, concluding remarks and future work are presented.

## METHODOLOGY AND FORMULATION

In this section, we detail the vehicle dynamic model, associated constraints, the proposed platooning controller as well as the settings to test the proposed logic.

### Vehicle Dynamic Constraints

Vehicles on the road are subject to various external forces and constraints. These include, dynamic forces, such as tractive and resistive forces, velocity, and acceleration constraints [13]. The tractive force is defined in Equation (1), the resistive force is the sum of the aero dynamic resistance $R_a$ (Equation (2)), rolling resistance $R_r$ (Equation (3)) and grade resistance $R_g$ (Equation (4)). Therefore, the upper bound for the acceleration is given by Equation (5) [14, 15].

$$F = \min\left(\frac{3600\eta_d P}{v}, m_{ta} g \mu\right) \tag{1}$$

$$R_a = \frac{\rho C_d C_h A_f v^2}{2} \tag{2}$$

$$R_r = mg C_{r0}(C_{r1} v + C_{r2}) \tag{3}$$

$$R_g = mgG \tag{4}$$

$$a_{max}(t) = \frac{F(t) - R_a(t) - R_r(t) - R_g(t)}{m} \tag{5}$$

The maximum deceleration a vehicle can experience is given by Equation (6)

$$a_{min} = -(G+1)g\,\mu\,b_e \tag{6}$$

Here the different quantities introduced in Equations (1)-(6) are summarized in Table 1.

**Table 1: Description of the various symbols used in the model equations.**

| Symbol | Description | Symbol | Description |
|--------|-------------|--------|-------------|
| $\eta_d$ | driveline efficiency (unitless) | $P$ | vehicle power (kW) |
| $m_{ta}$ | mass of the vehicle on the tractive axle (kg) | $g$ | gravitational acceleration (m/s$^2$) |
| $\mu$ | coefficient of road adhesion or the coefficient of friction (unitless) | $\rho$ | air density at sea level (kg/m$^3$); |
| $C_d$ | vehicle drag coefficient (unitless) | $C_h$ | altitude correction factor (unitless) |
| $A_f$ | vehicle frontal area (m$^2$) | $C_{r0}$ | rolling resistance constant that varies as a function of the pavement type and condition (unitless) |
| $C_{r1}$ | second rolling resistance constant (h/km); | $C_{r2}$ | third rolling resistance constant (unitless) |
| $m$ | total vehicle mass (kg) | $G$ | roadway grade (unitless) |
| $b_e$ | braking efficiency | | |

Here $a_{min}$ and $a_{max}$ are the absolute bounds for the acceleration of the vehicle. However, when there are other vehicles on the road collision avoidance is of utmost importance. Therefore, we introduce another constraint on the acceleration. This constraint has the exclusive role of decelerating the vehicle to a velocity that of the vehicle ahead of it while at the same time keeping adequate spacing. To avoid collision, the minimum deceleration is given by Equation (7)

$$a_{collision} = \frac{b_{kinematics}^2}{(b_{desired} + g\ Gr\ )} \tag{7}$$

where $b_{desired}$ is the desired deceleration level, and

$$b_{kinematics} = \frac{\left(v_n^2 - v_{n-1}^2 + \sqrt{(v_n^2 - v_{n-1}^2)^2}\right)}{4\left(x_{n-1} - x_n - s_j\right)} \tag{8}$$

where $v_n$ is the velocity of the current vehicle, $v_{n-1}$ is the velocity of the vehicle ahead of it, $x_n$ is the position of the current vehicle, $x_{n-1}$ is the position of the vehicle ahead, and $s_j$ is the spacing at jam conditions.

It is important to mention here that $b_{kinematics}$ is the deceleration level needed for the following vehicle to reduce its speed to that of the vehicle in front with the stopping distance being equal to the distance gap separating them. Therefore, the acceleration of any given vehicle needs to satisfy the following conditions given in Equation (9).

$$\begin{cases} a_n(t) \leq a_{max}(t) & \text{if } a_n > 0 \\ a_{min}(t) \leq a_n(t) \leq a_{collision}(t) & \text{if } a_n \leq 0 \end{cases} \tag{9}$$

The RPA car-following model[16] accounts for the constraints on the acceleration in the perspective of velocity. The velocity of the vehicle following another one needs to satisfy the condition presented in Equation (10) [17].

$$v_n(t + \Delta t) = min \begin{cases} v_n(t) + a_{max}(t)\,\Delta t \\ \dfrac{-c_1 + c_3 v_f + s_n(t + \Delta t) - \sqrt{A}}{2\,c_3} \\ \sqrt{v_{n-1}(t + \Delta t)^2 + 2\,b_{desired}\left(s_n(t + \Delta t) - \dfrac{1}{k_j}\right)} \end{cases} \tag{10}$$

Where $c_1 = \frac{v_f}{k_j v_c^2}(2\,v_c - v_f)$, $c_2 = \frac{v_f}{k_j v_c^2}(v - v_f)^2$, $c_3 = \frac{1}{q_c} - \frac{v_f}{k_j v_c^2}$, $s_n(t + \Delta t) = s_n(t) + [v_{n-1}(t) - v_n(t)]\Delta t + \frac{1}{2}a_{n-1}(t)\Delta t^2$, $A = [c_1 + c_3 v_f - s_n(t + \Delta t)]^2 - 4\,c_3[s_n(t + \Delta t)\,v_f - c_1 v_f - c_2]$, $v_f$ is the free flow velocity, $v_c$ is the velocity at capacity ($v_c \approx 0.85\ v_f$), $k_j$ is the jam density, and $q_c$ is the saturation flow rate.

It is important to mention here that the car following model presented in Equation (10) is enforced at all times throughout the simulation, particularly for vehicles not forming platoons.

## Proposed Controller

To maintain a constant time gap between two consecutive vehicles, the introduction of a controller is necessary. Provided feedback from various sensors, the objective of the controller is to maintain a constant/desired time gap $h_{des} = 0.6\ (s)$. This can be achieved by driving the error function – which transforms the desired time gap to a distance gap between two consecutive vehicles – defined in Equation (10) to zero. This can be achieved by letting the following vehicle apply successive and corrective acceleration/deceleration inputs. One of the simple ways of achieving this result is by enforcing that the time rate of change of the error $e_n(t)$ (i.e. $\frac{d}{dt}\big(e_n(t)\big)$) using Equation (11)

$$e_n(t) = \big[x_{n-1}(t) - x_n(t) - s_j\big] - h_{des} \times v_n(t) \qquad (11)$$

$$\frac{d}{dt}\big(e_n(t)\big) = -\lambda\, e_n(t) \qquad (12)$$

where $\lambda$ is a strictly positive real number. The solution to Equation (11) is given by

$$e_n(t) = e_n(0)\exp[-\lambda\, t]$$

and guaranties that $e_n(t)$ converges to zero as time increases provided $\lambda$ is strictly positive. Substituting Equation (10) into (11) leads to

$$a_n = \frac{-\lambda\, e_n(t) + v_{n-1} - v_n}{h_{des}} \qquad (13)$$

namely,

$$a_n(t) = \frac{1}{h_{des}}\big[-\lambda\left(x_{n-1}(t) - x_n(t) - s_j\right) + v_{n-1}(t) + (\lambda\, h_{des} - 1)v_n(t)\big] \qquad (14)$$

Equation (12) requires knowledge of the difference in position between two consecutive vehicles as well as their respective velocities, which can be achieved by having sensors on the vehicle or through V2V communication. The presented controller has one hyper-parameter. The amount of data that needs to be transferred between the vehicles is minimum (i.e. velocity of the vehicle ahead). It is also possible to avoid this transfer of information by estimating the position and velocity of the vehicle in front via radar.

It is also important to note that the computed value for the acceleration $a_n(t)$ need to satisfy conditions presented in Equation (9).

## Simulation Setup

In this paper we consider testing the introduced platooning logic on downtown Los Angeles, specifically, the highway stretches that traverse it from north to south and east to west. The total length selected for the platooning is approximately 123 km (~76 miles). The average free flow velocity, velocity at capacity, saturation flow rate and jam density are $v_f = 100$ km/h, $v_c = 85$ km/h, $q_c = 2480$ veh/h/lane, $k_j = 180$ veh/km/lane, respectively for the selected links. This implies a saturation headway of approximately 1.45 s. Consequently, the platooning produces a 45% reduction in the vehicle headways (a time gap of 0.6s corresponds to a headway of 0.8s considering a free-flow speed of 100 km/h). The selected area is presented in Figure 2. The network was modeled using the INTEGRATION software, which is developed by the Center of Sustainable Mobility (CSM) at the Virginia Tech Transportation Institute [18, 19]. The traffic demand was calibrated using loop detector data by computing the maximum likelihood static OD matrix using procedures described in [20] and then adjusting the static OD matrix to compute the dynamic OD matrix using procedures described in [21]. A detailed description of the calibration effort can be found in [22]. This resulted in a total of approximately 144,000 trips over a duration of 1 hour that was simulated to test the proposed algorithm.

One detail we encountered is that the selected highways do have different lane counts. This count ranges from 3 lanes to 6 lanes. For the purpose of this study, we select the two most left lanes as the lanes where we activate platooning. To further simplify the simulation, we assume that we have a single vehicle type. That is the 2018 Toyota Camry LE 2.5. The fleet of Toyotas are subdivided into two classes: class 1 and class 2. Class 1 are the Toyotas that do not form or join a platoon non-CACC equipped vehicles. Class 2 are the Toyotas that do form and if possible, join other created platoons (CACC-equipped vehicles). The ratio of Class 2 with respect to 1 was selected to be variable to discern the effects of various MPRs.

The amount of fuel consumed by this vehicle is given by Equation (13) [23], which is included in the INTEGRATION software.

$$FC = \alpha_0 + \alpha_1 P + \alpha_2 P^2 + \alpha_3 v \qquad (15)$$

where, $P$ is the power of the vehicle and $v$ is the velocity.

The power of the vehicle is the product of the force experienced by the vehicle and its velocity.

$$P = F_e \, v \qquad (16)$$

where,

$$F_e = mgG + \frac{mg \cos(\theta) \, CR_0 CR_2}{1000} + m \, a + \frac{mg \cos(\theta) \, CR_0 CR_1}{1000} \, v \qquad (17)$$
$$+ \frac{1}{2} \rho \, C_d \, C_h \, A_f \, v^2$$

where $a$ is the acceleration of the vehicle.

The delay that can be experienced by vehicles is computed using Equation (16).

$$delay = \sum_t \frac{v_f - v}{v_f} \, \Delta t \qquad (18)$$

where $v_f$ is the free flow velocity on a given link. This model was validated in [24].

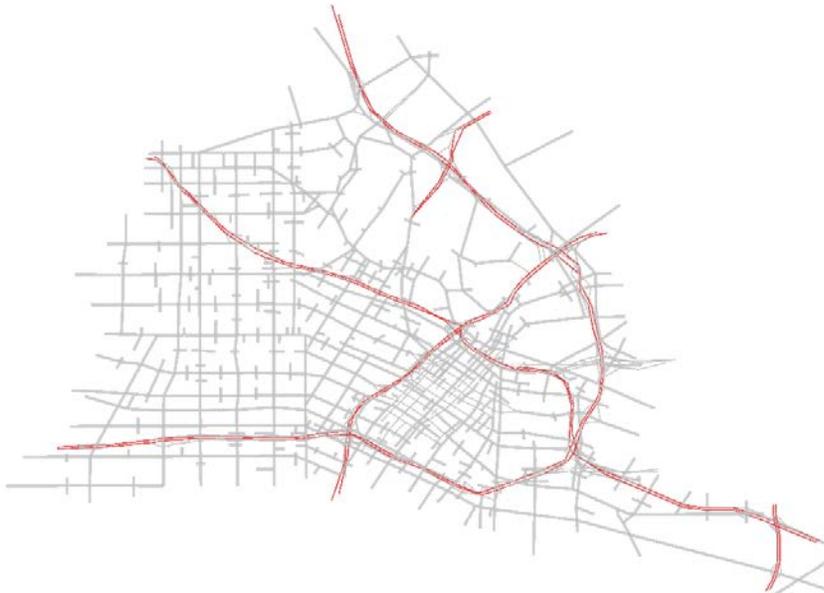

**Figure 1: Illustration of the downtown Los Angeles selected area for simulation. The roads highlighted in red are the selected highways over which platooning between vehicles is applied**

## RESULTS AND DISCUSSION

In order to account for varying traffic conditions from one day to the other, various simulations with different random seeds were performed. At first, we want to determine the best platooning configuration to adopt. A series of simulations are performed using the configurations listed in **Table 2**.

**Table 2: Tested Platooning Configurations**

| Configuration Name | Details |
|---|---|
| A | Platooning on all lanes of the highway |
| B | Platooning on all lanes of the highway – platoon size limited to 22 cars |
| C | Platooning on 1 lane |
| D | Platooning on 1 lane – platoon size limited to 22 cars |
| E | Platooning on 2 lanes – platoon size limited to 22 cars |

It is important to mention here that in certain configurations of Table 1 (B, D, and E) platooning is enforced on individual links in a disconnected manner from the links that follow. Noting that the average link length is 500 meters, the speed limit (i.e. platooning speed) is 27.78 m/s (100 km/h), and the selected time gap is $h_{des} = 0.6\ (s)$, a single vehicle occupies 22.67 meters. Therefore, 500 meters contain approximately 22 vehicles. Platoons are formed in a dynamic manner. Any vehicle attempting to join a platoon can increase its velocity by 7% beyond the speed limit (i.e. platooning speed) for a maximum duration of 6.5 (s). If the vehicle is unable to join the platoon within this time frame, a new platoon is formed with this vehicle as a lead vehicle. These parameters are user-specified and thus can be varied.

The average of the results for all the selected seeds are reported in **Table 3**. It is clear from **Table 3** that the best configuration leading to the best performance is configuration E. This corresponds to travel time, delay and fuel consumption reduction of 7.74 %, 13.6 % and 11.42% respectively. This configuration stipulates that platooning is enforced on the two leftmost lanes while limiting the size of the platoon. In the subsequent simulations, we will only consider configuration E.

In order to further investigate the effectiveness of platooning, we ran various simulations (different random seeds) with different MPRs. An average of the various results is reported in **Table 4**. **Table 4** shows the average travel time, delay and fuel consumed for all vehicles traversing the selected area for various MPRs. We can clearly discern that up to an MPR of 20 % no significant advantage is provided by the CACC platooning. In fact, the performance metrics are about the same. Starting from an MPR of 30%, we observe a reduction between 0 and 5% in travel time, a reduction between 0 and 9.4% in delay, and a reduction between 2.58 and 8.17% in fuel consumption. It is important to mention here that the RPA car-following model and collision avoidance [16] are enforced at all times between all the vehicles (platooned and non-platooned). The performance metric reported in **Table 2** are for all the vehicles inside the test area of downtown Los Angeles. This finding demonstrates that efficient movement of a subset of vehicles inside a large network leads to an improved mobility for the entire network.

**Table 3: Results for the various tested configuration listed in Table 2 for three different seeds(the negative percentages denote reductions).**

| | No Platooning | A | B | C | D | E |
|---|---|---|---|---|---|---|
| Travel Time (s) | 1032.60 | 1056.01 | 1085.74 | 993.30 | 1036.55 | 952.64 |
| Total Delay (s) | 561.70 | 566.21 | 587.20 | 513.40 | 568.03 | 485.30 |
| Fuel (l) | 0.89 | 0.74 | 0.84 | 0.88 | 0.88 | 0.79 |
| Travel Time Reduction (%) | | 2.27 | 5.15 | -3.81 | 0.38 | -7.74 |
| Delay Reduction (%) | | 0.80 | 4.54 | -8.60 | 1.13 | -13.60 |
| Fuel Consumption Reduction (%) | | -16.48 | -5.50 | -1.19 | -0.72 | -11.42 |

**Table 4: Summary of the simulation results inside downtown Los Angeles for various MPRs for six different seeds (the negative percentages denote reductions).**

| MPR (%) | 0% | 1% | 5% | 10% | 15% | 20% | 30% | 40% | 50% | 60% | 70% | 80% | 90% | 100% |
|---|---|---|---|---|---|---|---|---|---|---|---|---|---|---|
| Travel Time (s) | 986.2 | 1007.8 | 979.7 | 989.8 | 1005.1 | 1001.1 | 947.5 | 961.3 | 978.9 | 944.7 | 953.6 | 936.8 | 968.4 | 994.7 |
| Total Delay (s) | 519.5 | 538.2 | 511.7 | 524.8 | 536.0 | 530.6 | 481.7 | 497.2 | 513.9 | 481.8 | 488.8 | 470.7 | 497.9 | 522.8 |
| Fuel (l) | 0.8617 | 0.8727 | 0.8566 | 0.8631 | 0.8679 | 0.8632 | 0.8306 | 0.8357 | 0.8395 | 0.8131 | 0.8108 | 0.7913 | 0.8006 | 0.8087 |
| Travel Time Reduction (%) | | 2.19 | -0.67 | 0.36 | 1.92 | 1.50 | -3.93 | -2.52 | -0.75 | -4.21 | -3.31 | -5.02 | -1.81 | 0.85 |
| Delay Reduction (%) | | 3.60 | -1.50 | 1.02 | 3.19 | 2.14 | -7.27 | -4.28 | -1.07 | -7.25 | -5.91 | -9.40 | -4.16 | 0.64 |
| Fuel Consumption Reduction (%) | | 1.28 | -0.59 | 0.17 | 0.72 | 0.18 | -3.61 | -3.02 | -2.58 | -5.63 | -5.91 | -8.17 | -7.09 | -6.15 |

Figure 3, Figure 4, and Figure 5 present a scatter plot of the reduction in travel time, delay and fuel consumption reported in **Table 4**. Even though various seeds were used for the simulation, the plots stress the reduction in the mentioned performance metrics. The slope of the decline of the fuel consumption is steeper than the other measures of effectiveness. This is essentially due to the significant reduction in the aerodynamic force to which the platooned vehicles are subjected and therefore the vehicle needs less energy to overcome that force. The slope for the reduction in travel time, delay, and fuel consumption are approximately -3.5%, -6.9%, and -9% respectively. The respective coefficients of determination $R^2$ are 0.23, 0.28, and 0.87 which further stresses the steep reduction in fuel consumption due to platooning.

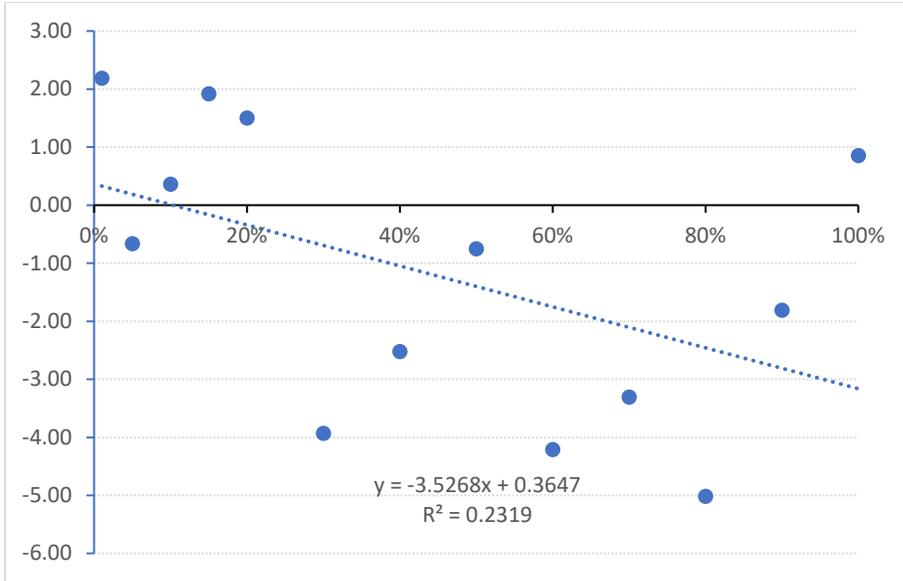

**Figure 2: Scatter plot of the travel time reduction due to platooning for all the vehicles inside the network as a function of the MPR.**

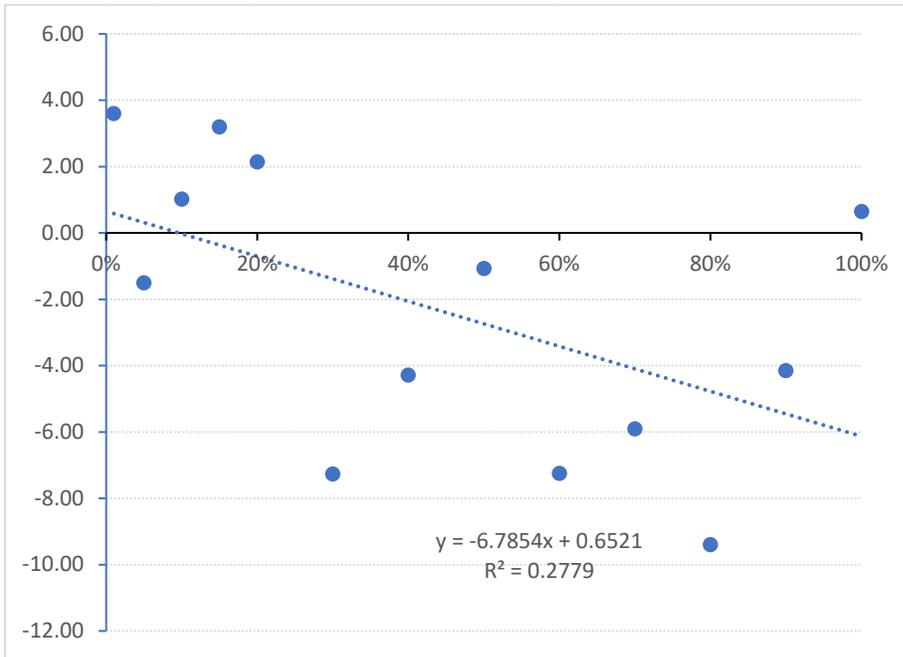

**Figure 3: Scatter plot of the delay reduction due to platooning for all the vehicles inside the network as a function of the MPR.**

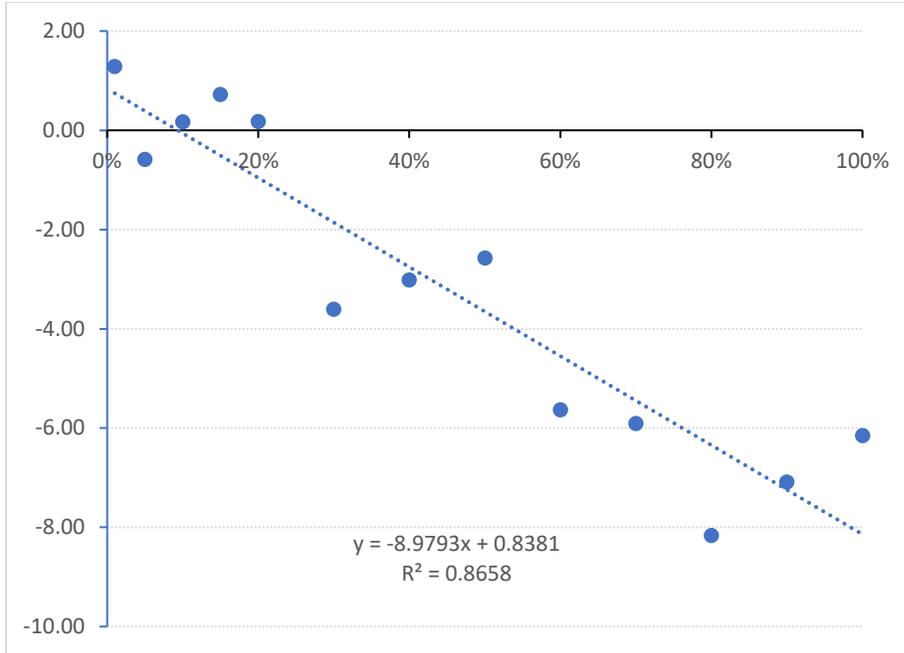

**Figure 4: Scatter plot of the fuel consumption reduction due to platooning for all the vehicles inside the network as a function of the MPR.**

## CONCLUSIONS

In this paper, we introduced a simple, input minimal, platooning logic. This logic takes into account various dynamic and kinematic constraints that vehicles experience. These include acceleration, velocity, and collision avoidance constraints. This control algorithm was later applied along the highways traversing downtown Los Angeles using the INTEGRATION software. The results suggest a clear trend towards a reduction in system-wide travel time, delay and notably fuel consumption. The average reduction in travel time for all the MPRs tested ranges from 0.75 to 5%. The average reduction in delay as well as fuel consumption (and ultimately $CO_2$ emission) ranges from 0 to 7% and 2.6 to 7% respectively. These results are for the fleet of all vehicles, platooned and non-platooned transiting through the downtown area. This leads us to deduce that controlling the trips of a subset of vehicles inside a large network does have the potential to impact other road users in a positive manner.

In future work, we will be conducting a detailed investigation on the performance of this controller on a mixed platoon comprised of conventional, hybrid and electrical vehicles. Various MPRs impact for the different vehicle types to be tested will be investigated using our in-house simulation software INTEGRATION.

## ACKNOWLEDGMENTS

This effort was funded through the Office of Energy Efficiency and Renewable Energy (EERE), Vehicle Technologies Office, Energy Efficient Mobility Systems Program under award number DE-EE0008209.

## AUTHOR CONTRIBUTIONS

The authors confirm contribution to the paper as follows: study conception and design: H.A. Rakha and Y. Bichiou; data collection: Y. Bichiou; analysis and interpretation of results: Y. Bichiou, H.A. Rakha, and H. Abdelghaffar; draft manuscript preparation: Y. Bichiou, H. Rakha, and H. Abdelghaffar. All authors reviewed the results and approved the final version of the manuscript.